\documentclass[aps,prl,twocolumn,showpacs,superscriptaddress]{revtex4}
\usepackage{graphicx}
\begin{document}

%\preprint{}
\title{Superconductivity coexisting with phase-separated static
magnetic order in (Ba,K)Fe$_{2}$As$_{2}$, (Sr,Na)Fe$_{2}$As$_{2}$ and
CaFe$_{2}$As$_{2}$\/}
     \author{T.~Goko}
     \affiliation{TRIUMF, 4004 Wesbrook Mall, Vancouver, B.C., V6T 2A3, Canada} 
     \affiliation{Department of Physics, Columbia University, New York, New York 10027, USA}
     \affiliation{Department of Physics and Astronomy, McMaster University, Hamilton, Ontario L8S 4M1, Canada}
     \author{A.~A.~Aczel}
     \affiliation{Department of Physics and Astronomy, McMaster University, Hamilton, Ontario L8S 4M1, Canada}
     \author{E.~Baggio-Saitovitch}
     \affiliation{Centro Brasilieiro de Pesquisas Fisicas, Rua Xavier Sigaud 150 Urca, CEP 22290-180
Rio de Janeiro, Brazil}
     \author{S.~L.~Bud'ko}
     \affiliation{Department of Physics and Astronomy and Ames Laboratory, Iowa State University, Ames, Iowa 50011, USA}     
     \author{P.C.~Canfield}
     \affiliation{Department of Physics and Astronomy and Ames Laboratory, Iowa State University, Ames, Iowa 50011, USA} 
     \author{J.~P.~Carlo} 
     \affiliation{Department of Physics, Columbia University, New York, New York 10027, USA}
     \author{G.~F.~Chen}
     \affiliation{Beijing National Laboratory for Condensed Matter Physics, Institute of Physics,
Chinese Academy of Sciences, Beijing 100080, Peoples Republic of China}
     \author{Pengcheng~Dai}
     \affiliation{Department of Physics and Astronomy, University of Tennessee, Knoxville, Tennessee 37996, USA}
     \author{A.~C.~Hamann}
     \affiliation{Forschungszentrum Karlsruhe, Institut f\"ur Festk\"orperphysik, Postfach 3640, D-76021
Karlsruhe, Germany}
     \author{W.~Z.~Hu}
     \affiliation{Beijing National Laboratory for Condensed Matter Physics, Institute of Physics,
Chinese Academy of Sciences, Beijing 100080, Peoples Republic of China}
    \author{H.~Kageyama}
    \affiliation{Department of Chemistry, Kyoto University, Kyoto 606-8502, Japan}
     \author{G.~M.~Luke}
     \affiliation{Department of Physics and Astronomy, McMaster University, Hamilton, Ontario L8S 4M1, Canada}
     \author{J.~L.~Luo}
     \affiliation{Beijing National Laboratory for Condensed Matter Physics, Institute of Physics,
Chinese Academy of Sciences, Beijing 100080, Peoples Republic of China}
     \author{B.~Nachumi}
     \affiliation{Department of Physics, Columbia University, New York, New York 10027, USA}
     \author{N.~Ni}
     \affiliation{Department of Physics and Astronomy and Ames Laboratory, Iowa State University, Ames, Iowa 50011, USA}  
     \author{D.~Reznik}
     \affiliation{Forschungszentrum Karlsruhe, Institut f\"ur Festk\"orperphysik, Postfach 3640, D-76021
Karlsruhe, Germany}
     \author{D.~R.~Sanchez-Candela}
     \affiliation{Centro Brasilieiro de Pesquisas Fisicas, Rua Xavier Sigaud 150 Urca, CEP 22290-180
Rio de Janeiro, Brazil}
     \author{A.~T.~Savici}
     \affiliation{Department of Physics and Astronomy, Johns Hopkins University, Baltimore, Maryland 21218, USA}
     \author{K.~J.~Sikes}
     \affiliation{Department of Physics, Columbia University, New York, New York 10027, USA}
     \author{N.~L.~Wang}
     \affiliation{Beijing National Laboratory for Condensed Matter Physics, Institute of Physics,
Chinese Academy of Sciences, Beijing 100080, Peoples Republic of China}
    \author{C.~R.~Wiebe}
    \affiliation{Department of Physics, Florida State University, Tallahassee, Florida 32310, USA}
     \author{T.~J.~Williams}
     \affiliation{Department of Physics and Astronomy, McMaster University, Hamilton, Ontario L8S 4M1, Canada}
     \author{T.~Yamamoto}
     \affiliation{Department of Chemistry, Kyoto University, Kyoto 606-8502, Japan}
     \author{W.~Yu}
     \affiliation{Department of Physics and Astronomy, McMaster University, Hamilton, Ontario L8S 4M1, Canada}
     \author{Y.~J.~Uemura}
     \altaffiliation[author to whom correspondences should be addressed: E-mail
tomo@lorentz.phys.columbia.edu]{}
     \affiliation{Department of Physics, Columbia University, New York, New York 10027, USA}
\date{\today}

\begin{abstract}
{\bf
The recent discovery and subsequent developments of FeAs-based superconductors 
have presented novel challenges and opportunities in 
the quest for superconducting mechanisms in correlated-electron systems.  
Central issues of ongoing studies include interplay between superconductivity and magnetism
as well as the nature of the pairing symmetry reflected in the superconducting energy gap.
In the cuprate and RE(O,F)FeAs (RE = rare earth) systems, 
the superconducting phase appears without being
accompanied by static magnetic order, except for
narrow phase-separated regions at the border of phase boundaries.  By muon spin
relaxation measurements on single crystal specimens, here we show that superconductivity in the AFe$_{2}$As$_{2}$ 
(A = Ca,Ba,Sr) systems, in both the cases of composition and pressure tunings,
coexists with a strong static magnetic order in a partial volume fraction.  
The superfluid response from the remaining paramagnetic volume fraction of 
(Ba$_{0.5}$K$_{0.5}$)Fe$_{2}$As$_{2}$ exhibits a nearly linear variation in T at low 
temperatures, suggesting an anisotropic energy gap with line nodes and/or multi-gap effects.\/}
\end{abstract}

%\keywords{}
\maketitle

%para 1

The announcement of superconductivity in La(O,F)FeAs ($T_{c} = 26$ K) in February 2008 \cite{laofeashosono}
triggered an unprecedented burst of research activities in FeAs-based superconductors and their
parent systems.  By now, superconductivity has been reported in systems with four different
crystal structures, i.e., the ``1111'' systems RE(O,F)FeAs with Rare Earth = La, Nd, Ce, etc. 
%\cite{xhchen,gfchen,zren,holedopedhhwen,highpressureoxygen}, 
\cite{oldrefs26},
the ``122'' systems
AFe$_{2}$As$_{2}$ (A = Ba, Sr, Ca) tuned by chemical substitutions 
%\cite{bakfe2as2,srkfe2as2} 
\cite{oldrefs78}
or application 
of hydrostatic pressure \cite{cafe2as2pres,lonzarichpressure}, the ``111'' Li$_{x}$FeAs \cite{lifeas}, and the ``011'' 
$\alpha$-FeSe systems 
%\cite{mkwufese,fesepressure}
\cite{oldrefs1213}.
Extensive measurements by neutron scattering \cite{daineutron,daiphase,bakfe2as2neutron},
Moessbauer effect \cite{klausscondmat,bafe2as2moess,srfe2as2moessmusr}
and muon spin relaxation ($\mu$SR) 
\cite{klausscondmat,oldrefs2225,drewcondmat,carlocondmat,aczelcondmat}
%\cite{klausscondmat,luetkenscondmat,drewcondmat,carlocondmat,khasanovcondmat,aczelcondmat}
have been performed to characterize magnetism and superfluid responses.
Notable results include the observations of collinear antiferromagnetic order in undoped parent compounds of
the 1111 and 122 systems \cite{daineutron,bafe2as2neutron,cafe2as2goldman}, 
hyperfine splitting of $^{57}$Fe Moessbauer spectra and $\mu$SR frequencies indicative of a static moment size
ranging between 0.3 - 0.8 Bohr magnetons per Fe \cite{bafe2as2moess,srfe2as2moessmusr,aczelcondmat},
static magnetism in lightly-doped non-superconducting
systems with possible incommensurate or stripe spin correlations \cite{carlocondmat}, and
nearly linear scaling between T$_{c}$ and the superfluid density
%\cite{luetkenscondmat,drewcondmat,carlocondmat,khasanovcondmat} 
\cite{oldrefs2225,drewcondmat,carlocondmat}
in the 1111 systems following
the trend found in cuprate and other exotic superconductors
\cite{oldrefs2729}. 
%\cite{uemuraprl89,uemuraprl91,yamazakiprize}.

%para 2 

Three systematic studies of magnetic phase diagrams of the 1111 systems
\cite{luetkensphase,daiphase,drewphase}, however, exhibit significantly different
behaviors.  As a function of increasing (O,F) substitution, La(O,F)FeAs shows
an abrupt and first-order like evolution from an antiferromagnetic to superconducting state,
Ce(O,F)FeAs shows nearly second-order like evolution, and Sm(O,F)FeAs exhibits phase-separated coexistence
of static magnetism and superconductivity in a small concentration
region around the phase boundary. Despite these differences, superconductivity
appears mostly in the region without static magnetic order in the 1111 systems, similarly to the case of the 
cuprates.
In contrast, very little has been reported on the 
phase diagrams of the 122 systems.  Recent powder neutron measurements on
(Ba,K)Fe$_{2}$As$_{2}$ \cite{bakfe2as2neutron} found a phase diagram similar to the one for Sm(O,F)FeAs
with coexisting long-range magnetic order and superconductivity near the phase boundary.  
Due to the volume-integrated character
of neutron instensity, however, information pertaining to the ordered volume fraction is missing in that work.
We have also reported $\mu$SR measurements on a single crystal of (Ba$_{0.55}$K$_{0.45}$)Fe$_{2}$As$_{2}$
\cite{aczelcondmat}
which found the co-existence of phase-separated static magnetic order and superconductivity.  
However, the superfluid density of this crystal was much lower than that in the corresponding 1111 systems with 
comparable T$_{c}$'s, which is suggestive of insufficient carrier doping. 
In the 122 systems, more definitive studies of magnetic phase diagrams can be expected
due to the availability of large single crystals \cite{canfieldbak,ninicafe2as2,wangsrk},
recent improvement of the growth method, and applicability of pressure-tuning free from randomness due to substitution.

%para 3

In this paper, we report muon spin relaxation ($\mu$SR) measurements of superconducting single crystals of 
(Ba$_{0.5}$K$_{0.5}$)Fe$_{2}$As$_{2}$ ($T_{c}\sim$ 37 K) and (Sr$_{0.5}$Na$_{0.5}$)Fe$_{2}$As$_{2}$ ($T_{c}\sim$ 35 K)
in ambient pressure,
and of CaFe$_{2}$As$_{2}$ in ambient and applied pressure $p$ up to $p$ = 10 kbar, performed at
TRIUMF in Vancouver, Canada.  The former two 
crystals, prepared at the Institute of Physics in Beijing using the FeAs flux method \cite{wangsrk} and
weighing $\sim$ 100 mg and 40 mg respectively, were mounted with their ab-planes
perpendicular to the muon beam.  Over 1 g of CaFe$_{2}$As$_{2}$ crystals (in more than 100 
pieces), prepared in Ames Lab. using the Sn flux method \cite{ninicafe2as2}, 
were mounted in a pressure cell having a sample space of 7 mm in diameter and 10-15 mm long.  
The crystals were aligned with their ab-planes perpendicular to the muon beam at the M9B channel, 
where the initial muon spin 
polarization is tuned to be perpendicular to the beam direction.   
$\mu$SR measurements were performed in zero field (ZF)
and weak transverse field (WTF) of $\sim$ 30 - 50 G to study magnetic ordering, and in transverse field (TF)
of 300 - 500 G to study superfluid density.  
A recent study on (Ca,Sr)RuO$_{3}$ and MnSi in applied pressure
\cite{nphysmnsi113} has demonstrated  $\mu$SR's unique capability of determining volume fractions of regions 
with and without static magnetic order in systems having phase separation.  
Details of the $\mu$SR methods and pressure measurement techniques can be found in 
refs. \cite{reviewrmp,reviewscot,saviciprb} and the on-line Supplementary Documents A and B.
%para 4

Figure 1 shows (a) the muon spin precession frequencies observed in ZF-$\mu$SR and (b) 
the paramagnetic volume fraction derived from WTF-$\mu$SR
measurements in the (Ba,K) and (Sr,Na) crystals.  The solid lines show the reported results in 
the undoped parent
compounds BaFe$_{2}$As$_{2}$ \cite{aczelcondmat} and SrFe$_{2}$As$_{2}$ \cite{srfe2as2moessmusr}.    
In both systems, static magnetism sets in at temperatures well above the superconducting $T_{c}$'s,
in a large volume fraction of $\sim$ 90 \%\ in the (Sr,Na) system and 50 \%\ in the (Ba,K) system.
We observed two different precession frequencies in a given system, presumably coming from two 
different muon sites, as was the case in BaFe$_{2}$As$_{2}$ \cite{aczelcondmat}.
The frequencies in the superconducting samples are reduced from 
the values in the undoped compounds only by 20-30 \%, indicating that static magnetic order with a significant
Fe moment size exists in the magnetically-ordered regions.  These results demonstrate phase separation 
between magnetically-ordered and paramagnetic volumes.   However, it is difficult to estimate the 
length scale of each region from the $\mu$SR results alone.  

%para 5

In TF-$\mu$SR, the precession signal from the paramagnetic volume fraction exhibits damping   
below $T_{c}$ due to an inhomogeneous field distribution in the flux vortex lattice.
The relaxation rate $\sigma$, obtained by fitting the spectra to a Gaussian
function $\exp(-\sigma^{2}t^{2}/2)$, is given by $\sigma \propto 1/\lambda^{2} \propto n_{s}/m^{*}$,
where $\lambda$ is the penetration depth, $n_{s}$ is the superconducting carrier density, and 
$m^{*}$ is the effective mass \cite{reviewrmp,reviewscot,saviciprb}.
An increase of $\sigma$ was observed in both the (Ba,K) and (Sr,Na) crystals below the 
superconducting $T_{c}$'s.
Since the statistical accuracy of the data is much better for the former system with the larger
paramagnetic volume fraction, here we present the results of (Ba$_{0.5}$K$_{0.5}$)Fe$_{2}$As$_{2}$
in TF = 500 G applied parallel to the c-axis 
in Fig. 2(a) and (b).  The temperature dependence of $\sigma$ in (a) is nearly linear with T,  
as demonstrated by the good agreement with the scaled data from a YBCO system \cite{reviewrmp}.
The observed behavior is distinctly
different from the case for an isotropic energy gap shown by the broken line representing 
a calculation for BCS s-wave coupling.
The observed temperature dependence may be attributed to (1) line nodes 
in an anisotropic energy gap, or (2)
widely different magnitudes of multiple isotropic gaps as seen in  
calculations \cite{benfatto,wenhc1} based on multiple bands.  An ARPES measurement \cite{dingarpes}
on (Ba,K)Fe$_{2}$As$_{2}$ reported evidence for multiple gaps.

% para 5.5 
  
The absolute value of $\sigma(T\rightarrow 0)$ is about a factor of 3 larger than that observed
in (Ba$_{0.55}$K$_{0.45}$)Fe$_{2}$As$_{2}$ in our previous measurements \cite{aczelcondmat}.  Given 
that $H_{c2}$ anisotropy is relatively low near $T_{c}$, between 3.5 and
2.5 for $ H <$ 14 T \cite{canfieldbak}, and
decreases for higher fields \cite{wanghc2},  we plot the present results 
in the 
$\sigma(T\rightarrow 0)$ versus $T_{c}$ plot of Fig. 2(b) without corrections
for single-crystal to polycrystal conversion \cite{russoprb}.
The resulting point from the 
present (Ba,K) crystal (red solid square symbol)
indicates that the present system has a superfluid density
close to those in the 1111 systems with comparable $T_{c}$'s, and that a sufficiently doped 122 FeAs
system follows the nearly linear relationship between $T_{c}$ and $n_{s}/m^{*}$ found in 
the cuprates and 1111 systems.  This result disagrees with the $H_{c1}$
measurements on (Ba,K)Fe$_{2}$As$_{2}$ \cite{wenhc1} which found the superfluid density 
to have a similar temperature
dependence to our results, but the absolute values differed by a factor of $\sim$ 4.
As experienced in early studies on various cuprate systems which reported widely 
scattered values of $H_{c1}$, the determination of $H_{c1}$ can be affected by flux pinning. 
This feature might explain the origin of the disagreement. 

%para 6

The superconducting state can also be obtained by 
applying hydrostatic pressure to the undoped parent compounds of the 122 systems
\cite{cafe2as2pres,lonzarichpressure}.  In particular, the CaFe$_{2}$As$_{2}$ system shows
superconductivity below $T \sim$ 10 K at relatively low pressures $p$ of 3-8 kbar, which are  
attainable using the available $\mu$SR piston-cylinder pressure cell.  
Note that this cell uses Daphne oil as the pressure mediator.  
We studied static magnetic order of CaFe$_{2}$As$_{2}$ at ambient pressure and
at $p$ = 3.9, 6.2 and 9.9 kbar by performing WTF-$\mu$SR with WTF = 50 G.
Solid symbols in 
Fig. 3(a) show the paramagnetic volume fraction, obtained after subtracting the 
contribution from the pressure cell in which the single crystal specimens were placed
with their c-axis oriented parallel to the beam direction.  Open circle symbols represent
additional results in ambient pressure obtained without the pressure cell.
Figure 3(a) demonstrates that static magnetic order sets in at temperatures well above the
superconducting $T_{c}$ in a partial volume fraction both at $p$ = 3.9 and 6.2 kbar.
The static magnetism disappears at $p$ = 9.9 kbar where the superconducting state no longer exists.  
Figure 3(b) shows the low temperature ($T \rightarrow 0$) values of the volume
fraction of the magnetically-ordered region (from WTF data)  
as well as the muon spin precession frequency
in ZF-$\mu$SR, which is proportional to the size of the ordered Fe moment.
We present the resulting pressure-temperature phase diagram of CaFe$_{2}$As$_{2}$
in Fig. 3(c).  The superconducting phase boundary in this figure 
is based on the reported resistivity results \cite{cafe2as2pres}. 
At this moment, we cannot determine whether the superconductivity
exists in the paramagnetic volume alone or in the entire sample volume below $T_{c}$.
In any case, Figs. 3(a)-(c)  
indicate that a rather strong magnetism exists in a substantial
volume fraction below T = 50 - 100 K which is well above the superconducting $T_{c}$,
as in the cases of the (Ba,K) and (Sr,Na) crystals in ambient pressure.

%newpara

Resistivity \cite{cafe2as2pres} and neutron \cite{cafe2as2neutron}
measurements in applied pressure, the former (the latter) using a liquid (He gas)
pressure mediator, have been reported 
on CaFe$_{2}$As$_{2}$ single crystals prepared by an identical method to
that used for the present specimens \cite{ninicafe2as2}. In resistivity studies, a sharp jump
was observed at T = 170 K at ambient pressure, corresponding to the 
first-order tetragonal-to-orthorhombic
structural phase transition below which magnetic order was detected both by
neutrons and muons.  With increasing pressure this feature broadens and ordering
moves towards lower temperatures, which is qualitatively consistent with the present
results in Fig. 3(c).  The resistivity anomaly becomes invisible above $p \sim$ 4 kbar,
and the neutron magnetic Bragg peak intensity at T = 50 K becomes nearly equal to the background level at
$p$ = 6.3 kbar (Fig. 1(c) in ref. \cite{cafe2as2neutron}), 
while $\mu$SR detected magnetic order continuing to exist at $p$ = 6.2 kbar, albeit in a 
partial volume fraction.
A rather sharp structural change detected by neutrons around $p$ = 3 kbar does not have
a corresponding signature in the $\mu$SR data except for an onset of phase separation.
At this moment, it is not clear whether these apparent differences are due to 
(a) different sensitivity of these three probes to magnetic order, as described in 
more detail in the on-line Supplementary Document C, or (b) to possibly different
degrees of the strain gradient between measurements with pressure instruments using liquid and gas mediators,
or to both (a) and (b).  Further experimental efforts are required for 
elucidating details of structural and magnetic behaviors, especially in the 
pressure region of $p$ = 4-8 kbar.
  
% para 8

In Figure 4, we plot the precession frequencies observed in ZF-$\mu$SR against
the Neel temperatures for various 1111 and 122 systems including undoped, doped,
and superconducting systems in ambient and applied pressure 
\cite{carlocondmat,aczelcondmat,srfe2as2moessmusr,ceof06unpublished}.
Since the Fe-muon hyperfine coupling constants are different only by within
10 \%\ between the 1111 and 122 systems \cite{aczelcondmat}, the vertical 
axis is nearly proportional to the static Fe moment size
(see on-line Supplemental Document D for more details).
Figure 4 indicates that static magnetism disappears when the system 
approaches the superconducting state in the 1111 systems, while significantly
stronger magnetism (with higher $T_{N}$ and larger moment size) remains in 
the 122 systems deep into the superconducting pressure region.
This may be related to the more three-dimensional
structure of the 122 systems which generally favours magnetic order.
Phase separation between volumes with and without static magnetism at
the border of a collinear antiferromagnetic state and a non-magnetic spin-gap
state was observed in an insulating J$_{1}$-J$_{2}$ spin system
Cu(Cl,Br)La(Nb,Ta)$_{2}$O$_{7}$ \cite{uemuraj1j2prl}.  Detailed features
of the observed $\mu$SR spectra in this system are remarkably 
similar to the results in the present 122 systems.  This similarity
suggests an important role of the J$_{1}$-J$_{2}$ frustration played in determining
the magnetic behavior of the FeAs-based systems.

%para 9

Regarding the pairing symmetry, available experimental results
on the 1111 and 122 systems are divided between those favoring
an isotropic nodeless gap \cite{dingarpes,hashimotomicrowave} and those supporting line
nodes \cite{oldrefs4849}. 
%\cite{stm,gapnodes}.  
Some theories for an extended 
s-wave pairing \cite{dopingdependentgap} propose evolution
from nodeless behavior to line-node behavior with progressive
carrier doping.  Multi-gap features can also result in T-dependence
different from the BCS s-wave curvature \cite{benfatto,wenhc1}.  
The present results in Fig. 2(a) represent
the first high-statistics $\mu$SR data of the thermodynamic behavior of
the superfluid density obtained in single crystal specimens of the
FeAs-based superconductors.  The difference between our previous
and present results on the two different (Ba,K) crystals, contrasted in 
Fig. 2(a), may be explained by their different doping levels corresponding to  
different superfluid densities.  It will be important to 
develop theories of superconducting pairing symmetry including the effect of
phase-separated, co-existing static magnetic regions.  
Figure 2(a) has established at least one definite
case which does not agree with a single isotropic energy gap.
The nearly linear relationship between $T_{c}$ and the superfluid density
(Fig. 2(b)) followed by cuprates, 1111 FeAs, 122 FeAs, and A$_{3}$C$_{60}$ 
systems suggests the existence of an underlying generic principle common to condensation mechanisms
of all these exotic superconductors.

We acknowledge financial support from
US NSF DMR-05-02706 and 08-06846 (Material World Network) at Columbia,
US NSF DMR-07-56568 at U. Tennessee Knoxville,
US NSF DMR-08-04173 and Florida State at Florida State U.,
US DOE under contract No. DE-AC02-07CH11358 at Ames,
NSERC and CIFAR (Canada) at McMaster, 
CNPq on MWN-CIAM program at CBPF (Brazil),
NSFC, CAS, and 973 project of MOST
(China) at IOP, Beijing, and US-Japan cooperative
program at Kyoto U. from JSPS (Japan).

\section{\bf On-line Supplementary Document A:  $\mu$SR time spectra in ZF, WTF and TF\/}

In this section, we briefly show basic features of $\mu$SR measurements in zero field (ZF), 
weak transverse field (WTF) and transverse field (TF), together with time spectra
obtained on single crystal specimens of (Ba$_{0.5}$K$_{0.5}$)Fe$_{2}$As$_{2}$ in
the present work.
Figure 5(a) shows the ZF-$\mu$SR spectra.  The appearance of oscillations below
$T\sim 60$ K indicates an onset of a static internal field from static Fe moments.
The amplitude of this signal is proportional to the fraction of muons landing in 
the region with static magnetic order.  The frequency is
proportional to the magnitude of the static internal field, and thus to the size
of the ordered Fe moments. In the case
of Fig. 5(a), the oscillation frequency of $\sim$ 20 MHz signal corresponds to an internal field of about 1.5 kG.
The damping of this oscillation is caused by a dephasing process due to 
an inhomogneity of the static internal field caused by 
short-range static spin correlations, or incommensurate magnetism, and/or possible 
effects of randomness and imperfections.  In general, fluctuating internal fields due to 
dynamic spin fluctuations
and/or spin waves can also cause damping of the oscillation through the
energy-dissipative $1/T_{1}$ processes.  Static and dynamic origins of the damping
can be distinguished by their different responses in $\mu$SR spectra to application of longitudinal
external fields.    Long-lived oscillation signals in ZF can be expected for
commensurate long-ranged static magnetism, as shown in Figs. 1(a) and 2(a) of ref. 
\cite{aczelcondmat} for the cases of the parent compounds of the 1111 and 122
FeAs systems.

When a weak, external transverse field (WTF) of typically 30-100 G is applied, 
muon spins in the paramagnetic or non-magnetic
environment exhibit slow coherent precession around the WTF, as shown in Fig. 5(b).
In magnetically-ordered regions of most magnetic systems, a static internal field from 
surrounding ordered moments at the muon site is much larger than the applied WTF.  As a result, those muon spins
cannot join in the WTF precession and the signal from the magntically-ordered region
usually disappears within about $t \sim 200$ ns in WTF measurements.  This is due to inhomogeneous broadening of 
the vector sums of the internal plus external fields.  The remaining slowly-oscillating
signal represents contributions from the other muons
in the paramagnetic environment.  As shown in Fig. 5(b), it is very straightforward
to derive the amplitude of this slow oscillation.  Figures 1(b) and 3(a) of the present
work and Figs. 1(c) and 3(c) of ref. \cite{aczelcondmat} show  
the amplitude of such slowly oscillating signals in WTF.  This procedure provides another method to 
estimate volume fractions of the ordered and paramagnetic regions, in addition to 
the ZF precession signal amplitude which represents the former.  The results from the WTF and ZF methods
usually agree well, as shown in Fig. 1(c) of ref. \cite{aczelcondmat}.

In type-II superconductors below $T_{c}$, the application of an external field 
above $H_{c1}$ leads to the 
formation of a flux vortex lattice inside the superconductor, which has
a well-defined field distribution \cite{reviewrmp}.  This field distribution 
causes additional damping of the precession signal in both WTF and TF-$\mu$SR measurements.
This feature can be seen in WTF spectra in Fig. 5(b) below $T_{c}$ = 37 K.  
The optimal value of TF to study the effect of the penetration depth and flux vortex lattice,
however, generally ranges between 100 - 3000 G, since it should be above $H_{c1}$
and well below $H_{c2}$, with the distance between the adjacent flux vortices significantly
smaller than the penetration depth.  In the present study, we chose to apply TF = 500 G,
and performed measurements in the field-cooling procedure.  Figure 5(c) shows the time
spectra of TF-$\mu$SR, from which we derived the relaxation rate $\sigma$ in Fig. 2(a).

\section{\bf On-line Supplementary Document B: high pressure instrumentation used in the present work\/}

A piston-cylinder-type pressure cell was used to generate 
hydrostatic pressures up to $\sim$ 10 kbar in the present work.
The cylinder is made of a non-magnetic alloy MP35N 
and has an inside diameter of 7 mm and a wall thickness of 8 mm.
Figure 6 shows some photos of this cell and a schematic drawing
of the $\mu$SR spectrometer used in high pressure measurements.
The sample space was filled 
with a pressure-transmitting medium of Daphne oil 7373.
The pressure cell was pressurized at room temperature 
before it was mounted in the cryostat and the pressure was maintained by using CuBe screws. 
The pressure was generally found to decease by approximately 3 kbar at low temperatures 
due to a thermal contraction difference 
between the cell and the pressure-transmitting medium. 
In this paper, we used the low-temperatures-range pressure
which was calibrated from the pressure dependence 
of superconducting temperature in Pb.

Previous $\mu$SR measurements on MnSi \cite{nphysmnsi113}, performed
using this instrument at 0 - 16 kbar, reproduced the magnetic ordering temperatures and
sharp features at phase boundaries reported in prior
measurements by other techniques using different
pressure cells.  Sharp variations of $\mu$SR results have also
been observed in our unpublished $\mu$SR work on UGe$_{2}$.
Although these observations in other magnetic systems may or
may not be generalized to the case of CaFe$_{2}$As$_{2}$
which exhibits a large change of the unit cell volume
in the pressure range of the present work \cite{cafe2as2neutron},
the reproducibility and sharp aspects assure that 
the pressure cell used in the present work is as good, uniform and accurate as
typical piston-cylinder cells currently used in 
various high pressure experiments in terms of homogeneity
and absolute values of the generated pressure. 

\section{\bf On-line Supplementary Document C: Different sensitivity of muons 
and neutrons in detecting magentic order\/}

In magnetic systems that exhibit commensurate long-range, static magnetic order,
neutron scattering observes magnetic Bragg peaks, while $\mu$SR often finds
long-lived coherent precession in ZF below the ordering temperature.  This
was the case for LaOFeAs, as published in refs. \cite{daineutron,carlocondmat}
which report neutron and $\mu$SR measurements performed using an identical piece
of sintered polycrystalline specimen.  Neutron measurements are generally limited by small intensities
of magnetic Bragg peaks, especially for the case of polycrystalline specimens
having moments smaller than $\sim$ 0.5 Bohr magnetons per unit cell.
In contrast, $\mu$SR has a superb sensitivity for detecting small or random static moments, as even nuclear
dipolar moments can be easily detected.  In NdOFeAs, neutron studies initially 
reported the absence of static magnetic order, while $\mu$SR measurements found 
a static magnetic order very similar to that of LaOFeAs \cite{aczelcondmat}.  The magnetic 
Bragg peaks in this material 
were subsequently observed after repeated measurements with 
higher statistics \cite{neutronndofeas}.  This incident demonstrates the potential difficulty in
detecting static magnetic order by neutron scattering, even for the case of systems with 
long-range spin correlations.

For static magnetic order involving short-range or incommensurate or stripe spin
correlations, it becomes even more difficult for neutron measurements to detect
the onset of ordering, since the intensity is 
spread in a wider region of momentum space.  As an example, here we will compare
neutron and $\mu$SR results obtained in a ceramic polycrystalline specimen of 
Ce(O$_{0.94}$F$_{0.06}$)FeAs.  This system is not superconducting, but lies in the 
border region between the magnetically-ordered and suprconducting phases, as elucidated
by neutron scattering studies \cite{daiphase}.  The Bragg peak intensity for this 6\%\
F doped system was very low, and static order was detected only below $\sim$ 10 K.
In contrast, $\mu$SR detected static order below T $\sim$ 40 K
as shown by the oscillation amplitude in WTF-$\mu$SR in Fig. 7(a).  The asymmetry
of $\sim$ 0.2 at high temperatures corresponds to the value expected for fully 
paramagnetic systems, while the loss of the asymmetry below 40 K indicates
a sharp onset of static magnetic order in 100 \%\ of the volume fraction. The frequency 
of the muon precession in ZF-$\mu$SR (Fig. 7(b)) shows that the ordered moment
size is about 0.1 Bohr magnetons.  

This apparent difference of ordering behavior
found by neutrons and muons can be understood when we look at the spectra
of ZF-$\mu$SR in Fig. 7(c).  The spectra at T = 5K exhibits a fast damping corresponding to a 
very short-ranged character of static Fe spin correlations. The faster 
damping seen at T = 2 K is most-likely caused by imminent ordering of Ce.  
These examples on NdOFeAs and Ce(O,F)FeAs demonstrate that different magnetic behaviors found by
neutrons and muons can be caused by different sensitivities of these probes to 
static magnetic order.  

This general argument provides a possible explanation for 
the static magnetic order of CaFe$_{2}$As$_{2}$ found in a wider pressure region
in the present $\mu$SR measurements
as compared to that found by neutron scattering studies \cite{cafe2as2neutron}.
An increasingly short-ranged spin correlation is found with 
increasing pressure by the present studies of CaFe$_{2}$As$_{2}$,
as demonstrated by the ZF-$\mu$SR spectra in Fig. 7(d).

\section{\bf On-line Supplementary Document D: Conversion between the ZF-$\mu$SR
frequency and ordered Fe moment size\/}

Frequencies observed in ZF-$\mu$SR measurements are generally proportional to the size of the static ordered moment
within a given series of magnetic materials having the same crystal structures and an identical muon site location in the 
unit cell.  However, the conversion from the frequency to the Fe moment size is not easy in FeAs-based systems,
since the accurate location of the muon site is not known.  The argument in Supplementary Doc. C indicates
that neutron scattering intensity is not much reliable for this purpose either.
In contrast, the hyperfine splitting observed in Moessbauer effect measurements can provides a more
reliable estimate of the size of the static Fe moments.
As described in ref. \cite{aczelcondmat}, comparisons of $\mu$SR frequencies and Moessbauer
hyperfine fields observed in LaOFeAs, BaFe$_{2}$As$_{2}$ and SrFe$_{2}$As$_{2}$ 
\cite{klausscondmat,carlocondmat,aczelcondmat,bafe2as2moess,srfe2as2moessmusr}
indicate that the hyperfine coupling constants between the muon spin and the Fe moment
are nearly equal for the 1111 and 122 systems, with a difference of about 10 \%.

Therefore, we plot the $\mu$SR freqencies in the vertical axis of Fig. 4 without corrections
between the 1111 and 122 systems.  The absolute value of the moment was scaled based on 
the Moessbauer results in ref. \cite{bafe2as2moess} for BaFe$_{2}$As$_{2}$.  
The resulting Fe moment size, shown in the right-vertical axis of Fig. 4, involves about 10 \%\
error due to the difference between the 1111 and 122 systems.  
The results of this conversion agree rather well with the estimates from neutron scattering studies
which reported 0.36 and 0.8 Bohr magnetons for Fe moments in LaOFeAs \cite{daineutron}
and CaFe$_{2}$As$_{2}$ \cite{cafe2as2goldman} respectively.  
Different Moessbauer measurements reported in the literature have adopted different conversion 
factors between the hyperfine splitting and the Fe moment size
\cite{klausscondmat,bafe2as2moess}, resulting 
in estimates for the ordered moment size in LaOFeAs
that differ in magnitude up to 40 \%.  
This ambiguity leads to a 
systematic error for the absolute moment values in Fig. 4.
The relative comparison of the moment size between different systems, however,
can be made with the accuracy of +/- 10 \%, assuming that the muon site does not change
among materials within each series (1111 and 122).   
\\

\onecolumngrid
\vfill \eject
\newpage
\begin{figure}[p]
\includegraphics[width=4.5in,angle=0]{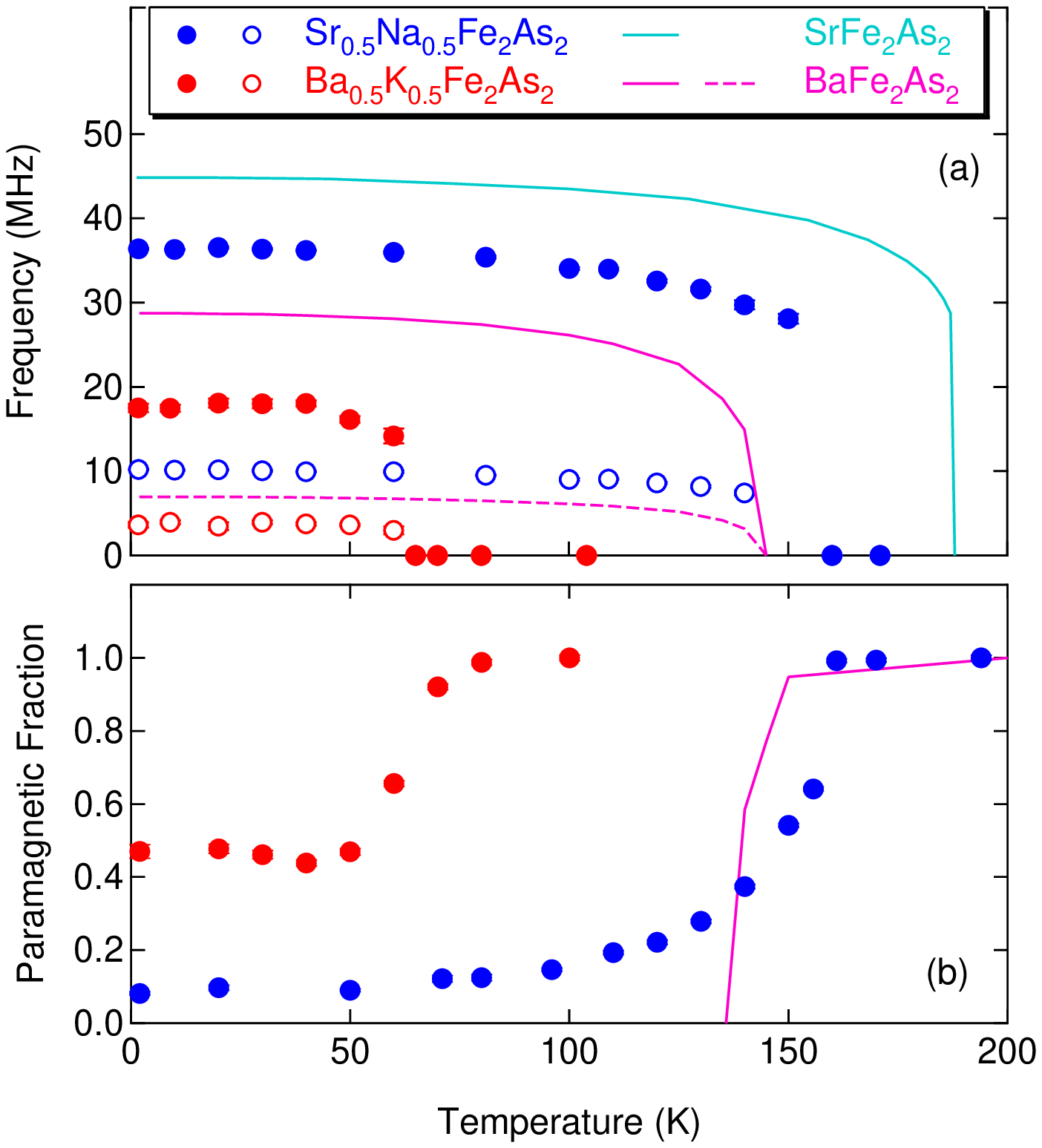}%
\caption{\label{fig1}(color)
Temperature dependences of (a) the muon spin precession frequency observed in ZF-$\mu$SR and
(b) the paramagnetic volume fraction determined in WTF-$\mu$SR measurements in 
single crystal specimens of (Ba$_{0.5}$K$_{0.5}$)Fe$_{2}$As$_{2}$ ($T_{c}\sim$ 37 K) and 
(Sr$_{0.5}$Na$_{0.5}$)Fe$_{2}$As$_{2}$ ($T_{c}\sim$ 35 K), which demonstrate the onset of 
static magnetic order in a partial volume fraction at temperatures well above the superconducting $T_{c}$. 
The results from the present work
(solid and open circles) are compared with those of the undoped parent compounds BaFe$_{2}$As$_{2}$
\cite{aczelcondmat} and SrFe$_{2}$As$_{2}$ \cite{srfe2as2moessmusr} (solid and broken lines).}
\end{figure}
%\onecolumngrid
\begin{figure}[p]
\includegraphics[width=4.0in,angle=0]{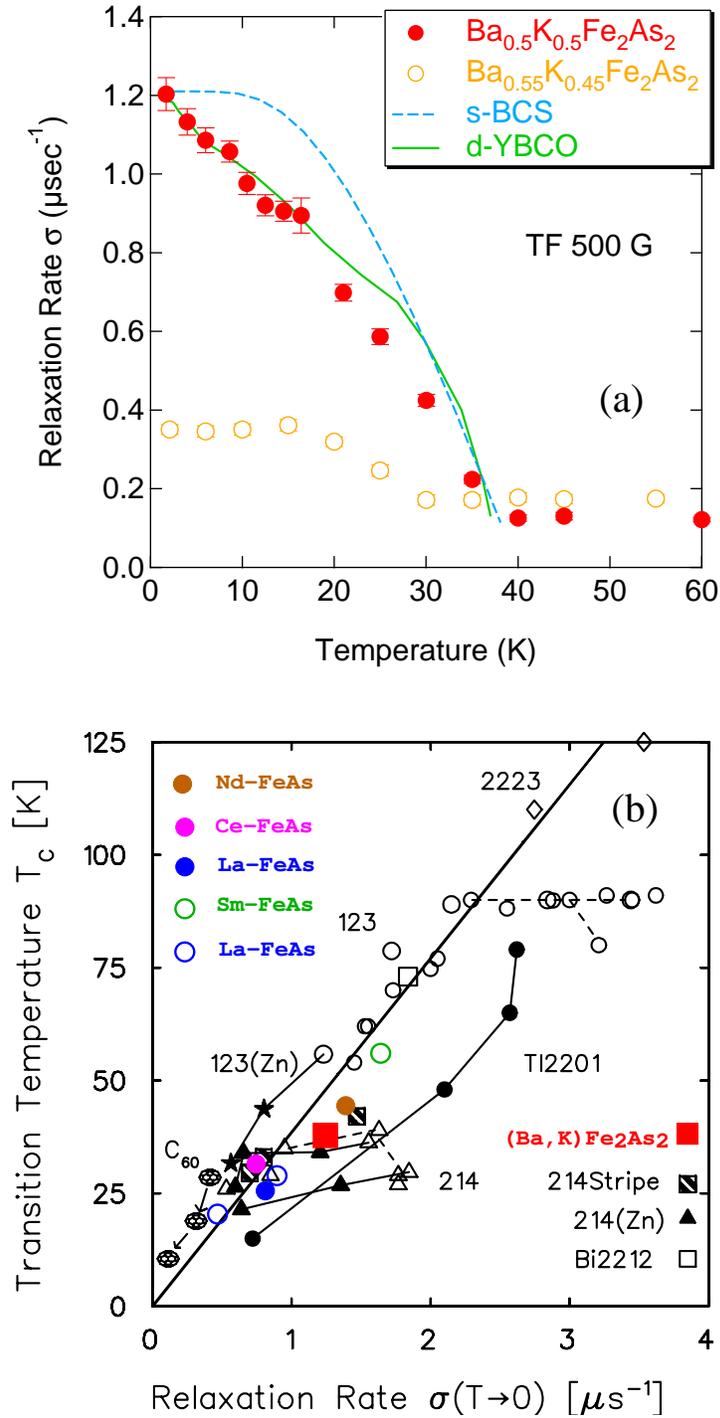}%
\caption{\label{fig2}(color)
(a) Temperature dependence of the muon spin relaxation rate $\sigma$ observed in a
single crystal specimen of (Ba$_{0.5}$K$_{0.5}$)Fe$_{2}$As$_{2}$ ($T_{c}\sim$ 37 K)
in TF-$\mu$SR with TF = 500 G (closed circles, present work),
compared with the temperature dependence expected for the isotropic energy gap of
BCS s-wave pairing (broken line), scaled results from YBCO \cite{reviewrmp}
(solid line), and our previous results in a different crystal of 
Ba$_{0.55}$K$_{0.45}$Fe$_{2}$As$_{2}$ \cite{aczelcondmat} (open circles).
The superfluid density is observed to be linear in T at low temperatures, which is consistent with an anisotropic
energy gap with line nodes or multiple isotropic gaps with widely different magnitudes. 
(b) A plot of the relaxation rate $\sigma(T\rightarrow 0)$ versus $T_{c}$, including
the point for (Ba$_{0.5}$K$_{0.5}$)Fe$_{2}$As$_{2}$ from the present work as well as
those for the FeAs-based 1111 superconductors published in refs. 
%\cite{luetkenscondmat,carlocondmat,khasanovcondmat},
\cite{oldrefs2225,carlocondmat},
various cuprates, and A$_{3}$C$_{60}$ superconductors 
\cite{oldrefs2729},
%\cite{uemuraprl89,uemuraprl91,yamazakiprize},
which demonstrates a nearly linear relationship between the superfluid density $n_{s}/m^{*}$
and $T_{c}$ followed by all of these exotic superconductors.}
\end{figure}
\newpage
\begin{figure}[t]
\includegraphics[width=3.7in,angle=0]{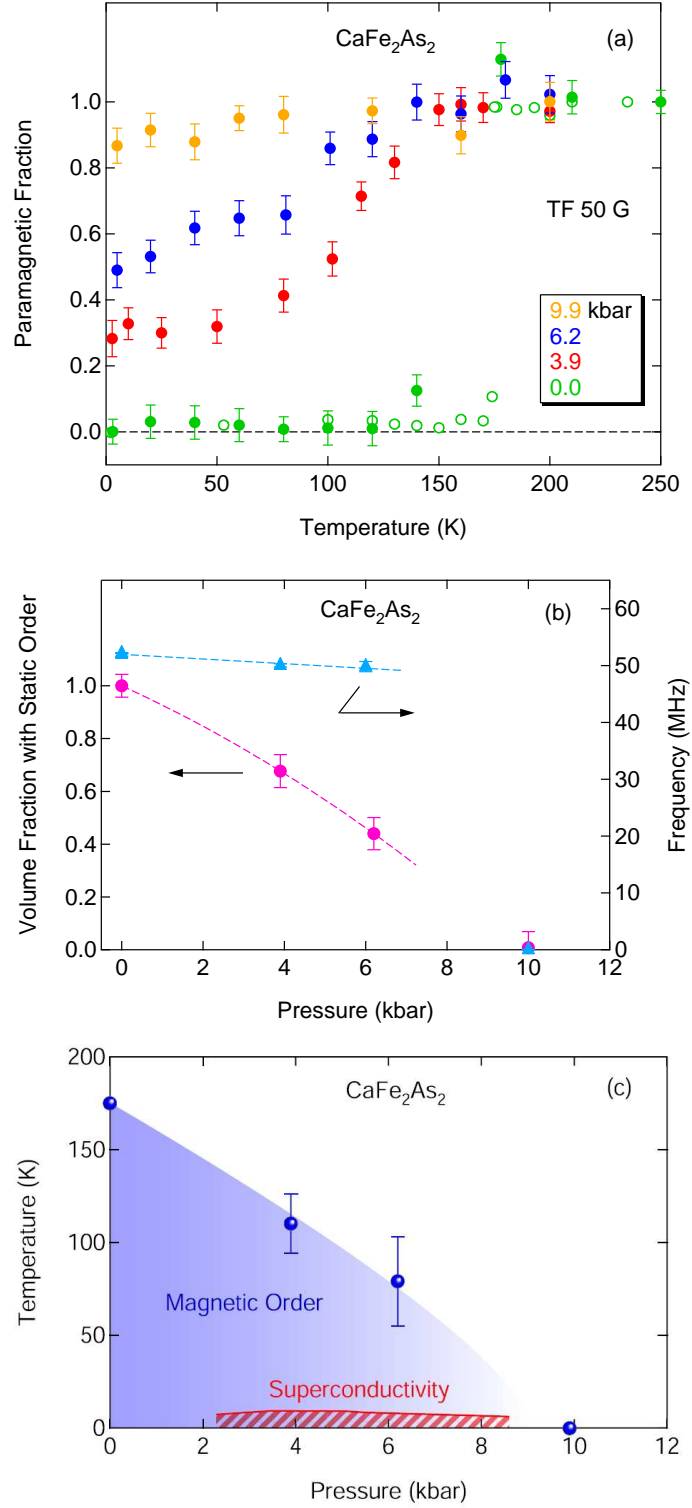}%
\caption{\label{fig3}(color)
(a) The volume fraction of regions without static magnetic order in CaFe$_{2}$As$_{2}$, 
as a function of temperature and pressure, determined by WTF-$\mu$SR measurements 
with WTF $\sim$ 50 G .  The points with closed (open) symbols were obtained in measurements with (without)
a pressure cell.  
(b) Pressure dependence of the volume fraction of the magnetically-ordered region (purple
closed circles; left axis) from WTF-$\mu$SR and the muon spin precession frequency 
(blue triangles; right axis) from ZF-$\mu$SR at $T \sim$ 2K in CaFe$_{2}$As$_{2}$.
(c) The phase diagram as a function of pressure and temperature in CaFe$_{2}$As$_{2}$.
These results demonstrate an onset of static magnetic order in a partial volume
fraction at temperatures well above the superconducting $T_{c}$'s.  The $T_{c}$ values are
taken from the reported resistivity results \cite{cafe2as2pres}.  Upper, middle, and lower temperatures
attached to the closed circle symbols for $T_{N}$ represent temperatures at which the volume fraction with
static magnetic order becomes 30, 50, and 70 \%\ of the value at T $\rightarrow$ 0. respectively}
\end{figure}
\vfill \eject
\begin{figure}[t]
\includegraphics[width=6.0in,angle=0]{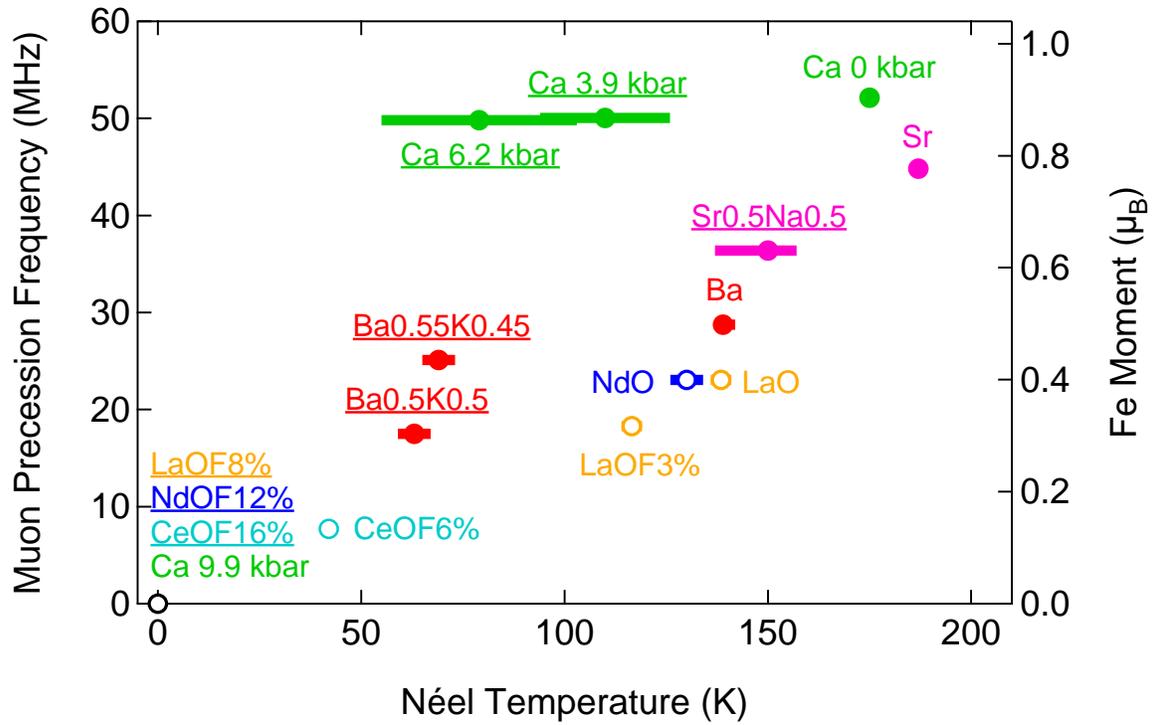}%
\caption{\label{fig4}(color)
Plot of the muon spin precession frequency in ZF-$\mu$SR versus Neel temperature
for the 1111 FeAs systems (open symbols; from refs. \cite{carlocondmat,ceof06unpublished}
and the 122 FeAs systems (filled symbols; present work and refs. \cite{aczelcondmat,srfe2as2moessmusr}).
Superconducting compounds are underlined.  An estimate of the ordered Fe moment size, based on 
comparisons of $\mu$SR and Moessbauer effect results \cite{aczelcondmat}, is given in the right axis.
Details of the frequency to moment conversion is given in the online Supplementary Document D.
The extended bar symbols for the superconducting 122 systems indicate temperatures
at which the volume fraction with static magnetic order becomes 30 and 70\%\ of the values at T = 2K.
These results demonstrate that the static magnetism is more robust in the 122 systems, which 
have more three-dimensional crystal structures than the 1111 systems.}
\end{figure}
\vfill \eject
\newpage
\begin{figure}[t]
\includegraphics[width=7.0in,angle=0]{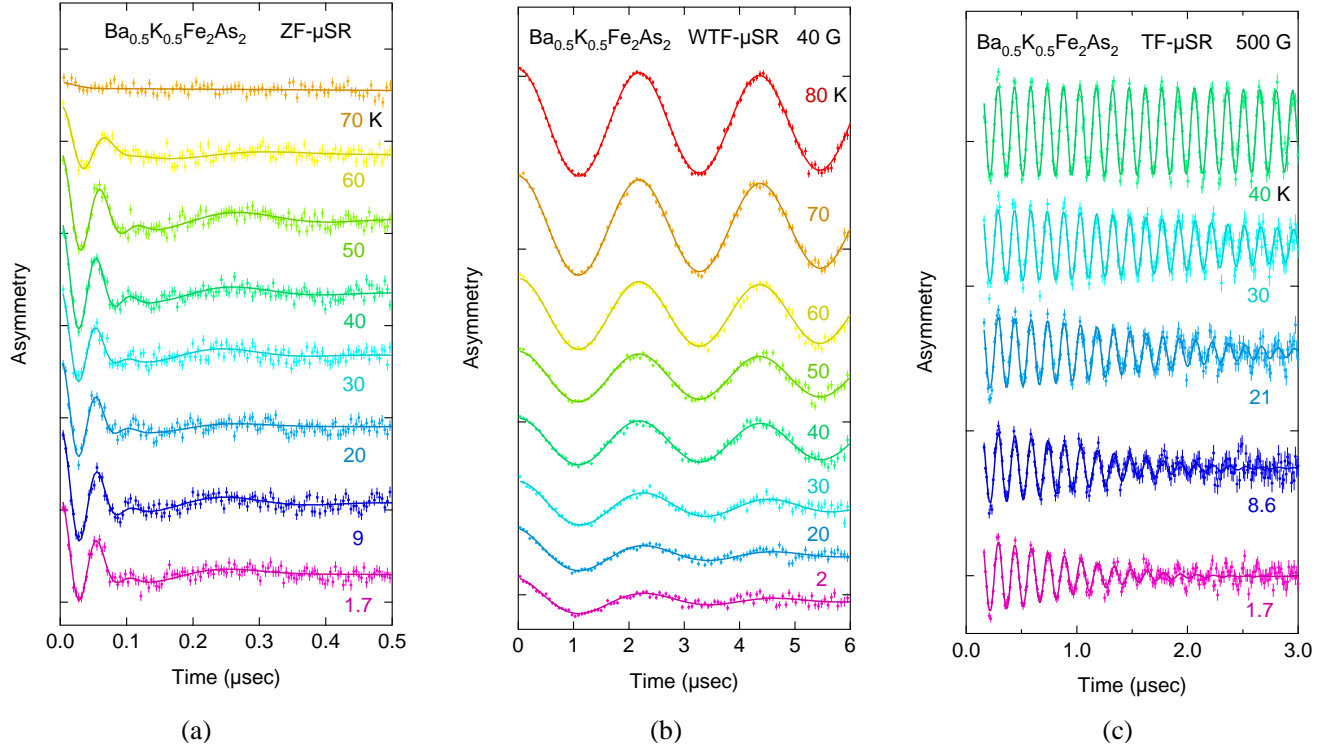}%
\caption{\label{fig5}(color)
{\bf Figure for the on-line Supplementary Document A\/}\\
Time spectra of (a) zero-field $\mu$SR, (b) weak-transverse-field $\mu$SR, and 
(c) transverse-field $\mu$SR observed in a single-crystal specimen of 
Ba$_{0.5}$K$_{0.5}$Fe$_{2}$As$_{2}$.}
\end{figure}
\newpage
\begin{figure}[t]
\includegraphics[width=5.0in,angle=0]{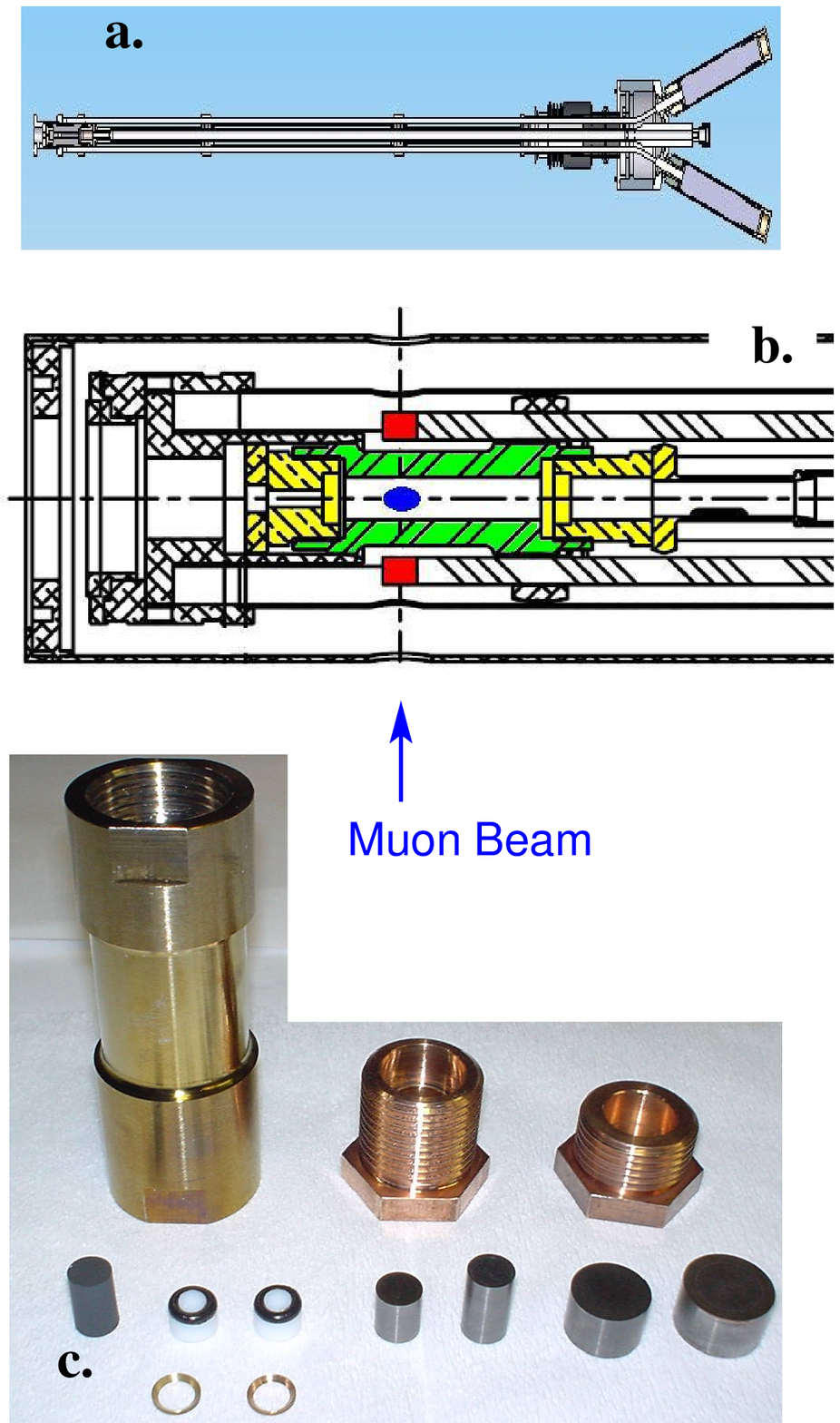}%
\caption{\label{fig6}(color)
{\bf Figure for the on-line Supplementary Document B\/}\\
(a) and (b): Schematic drawing of a $\mu$SR spectrometer at TRIUMF used for 
measurements with applied pressure.  The blue circle in (b) represents the 
specimen, and the green and yellow regions show the walls of the pressure cell.
The red squares show plastic scintillation counters which detect incoming muons
and decay positrons.  (c) A photograph of the piston-cylinder-type pressure cell
used in the present measurements.}
\end{figure}
\newpage
\begin{figure}[t]
\includegraphics[width=6.5in,angle=0]{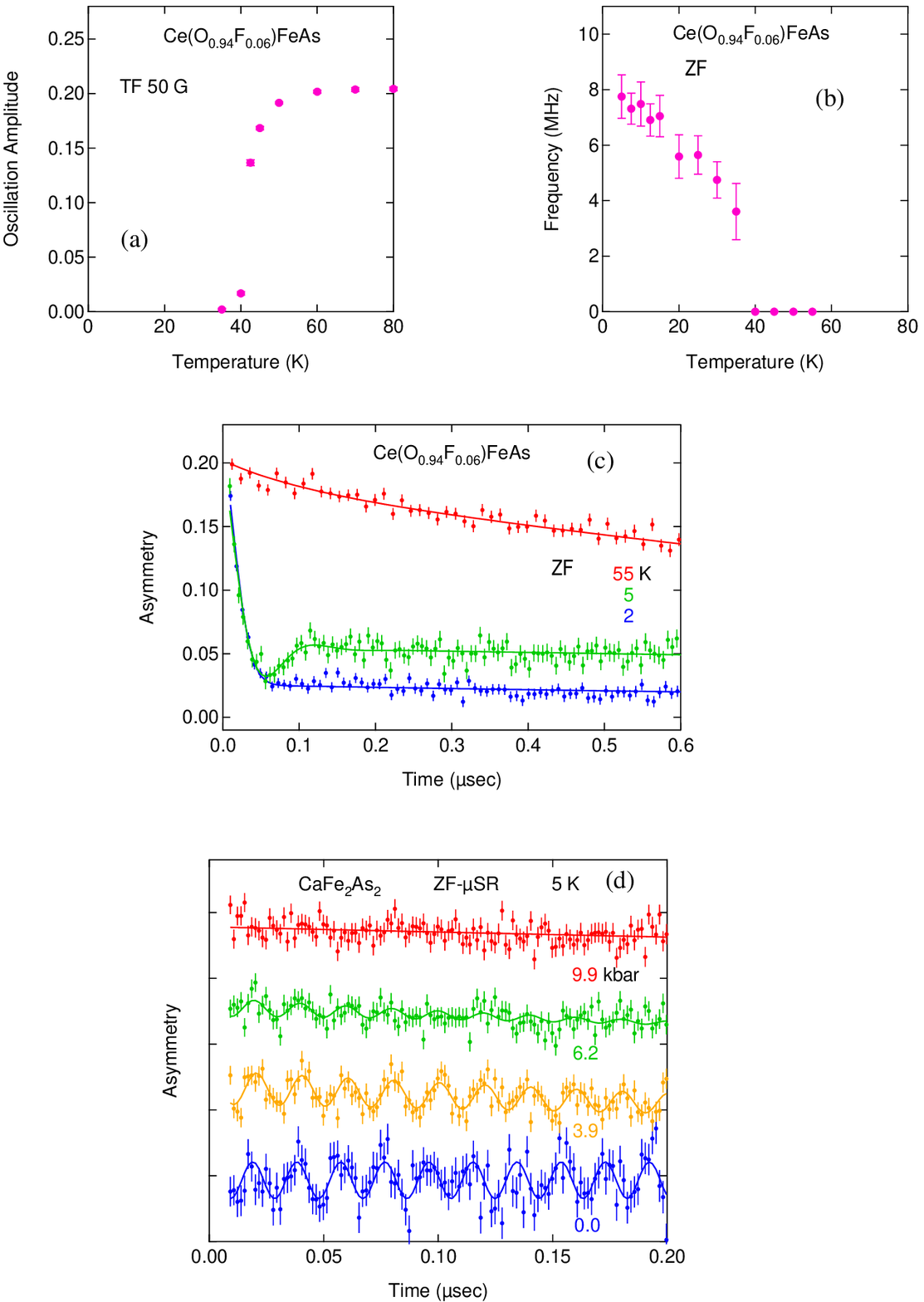}%
\caption{\label{fig7}(color)
{\bf Figure for the on-line Supplementary Document C\/}\\
(a) The precessing amplitude of muon spins in a weak transverse field (WTF) of 50 G,
(b) the ZF-$\mu$SR precession frequency, and (c) the time spectra of ZF-$\mu$SR
observed in a polycrystalline specimen of CeO$_{0.94}$F$_{0.06}$FeAs.
(d) shows the time spectra of ZF-$\mu$SR measurements from a mosaic single-crystal
specimen of CaFe$_{2}$As$_{2}$ obtained using a pressure cell at T = 5 K at several
different hydrostatic pressures.}
\end{figure}

\end{document}